\documentclass[notitlepage,aps,prl,reprint, groupedaddress]{revtex4-2}

\usepackage{mathtools}
\usepackage[english]{babel}
\usepackage[utf8]{inputenc}
\usepackage{amsmath}
\usepackage{amsfonts}
\usepackage{amssymb}
\usepackage{graphicx}
\usepackage{bbold}
\usepackage{hyperref}

\DeclareMathOperator{\sech}{sech}
\DeclareMathOperator{\Tr}{Tr}
\DeclarePairedDelimiter\bra{\langle}{\rvert}
\DeclarePairedDelimiter\ket{\lvert}{\rangle}

\begin{document}
\renewcommand{\figurename}{FIG.}
\renewcommand{\tablename}{TABLE}
\title{Analysis of loss correction with the Gottesman-Kitaev-Preskill code}
\author{Jacob Hastrup}
\email{jhast@fysik.dtu.dk}
\author{Ulrik Lund Andersen}
\email{ulrik.andersen@fysik.dtu.dk}

\affiliation{Center for Macroscopic Quantum States (bigQ), Department of Physics, Technical University of Denmark, Building 307, Fysikvej, 2800 Kgs. Lyngby, Denmark}

\begin{abstract}
The Gottesman-Kitaev-Preskill (GKP) code is a promising bosonic quantum error-correcting code, encoding logical qubits into a bosonic mode in such a way that many physically relevant noise types can be corrected effectively. A particularly relevant noise channel is the pure loss channel, which the GKP code is known to protect against. In particular, it is commonly pointed out that losses can be corrected by the GKP code by transforming the losses into random Gaussian displacements through a quantum-limited amplification channel. However, implementing such amplification in practice is not ideal and could easily introduce an additional overhead of noise from associated experimental imperfections. Here, we analyse the performance of teleportation-based GKP error correction against loss in the absence of an amplification channel. We show that amplification is not required to perform GKP error correction, and that performing amplification actually worsens the performance for practically relevant parameter regimes. 
\end{abstract}
\date{\today}
\maketitle

\subsection{Introduction}
Efficient quantum error correction (QEC) is essential to scale up quantum computers. One promising QEC technique that has gained increased interest in recent years is bosonic QEC, in which each qubit is encoded into a harmonic oscillator \cite{gottesman2001encoding,albert2018performance,grimsmo2020quantum,terhal2020towards,cai2021bosonic}. This allows for an error-correctable logical qubit to be defined within a single physical mode, providing a hardware-efficient approach to fault-tolerant quantum computing. The Gottesman-Kitaev-Preskill (GKP) code \cite{gottesman2001encoding,grimsmo2021quantum} is a bosonic code which is particularly relevant for optical systems \cite{tzitrin2020progress,fukui2021building}, as gates, error-correction and measurements can all be carried out efficiently with Gaussian interactions. Although it is a daunting task to generate the non-Gaussian GKP basis states on an optical platform, it is likely that they will become available in the future, e.g.\ using one of the numerous proposed protocols \cite{vasconcelos2010all,weigand2018generating,motes2017encoding, eaton2019non,su2019conversion,tzitrin2020progress,fukui2021efficient,hastrup2021generation,fukui2021generating}. 

The GKP code is designed to correct small phase-space displacement errors, and since any channel can be expanded in terms of such displacements \cite{gottesman2001encoding} the GKP code is in principle capable of correcting many different types of errors, provided the expansion only contains small displacements. Unfortunately, physically relevant channels, such as loss and dephasing, are generally not expanded only in terms of small displacements \cite{terhal2016encoding}. For example, the loss channel with transmissivity $\eta$ acting on a coherent state reduces the amplitude as $\alpha\rightarrow \sqrt{\eta}\alpha = \alpha - (1 - \sqrt{\eta})\alpha$, corresponding to a displacement of magnitude $(1 -\sqrt{\eta})\alpha$. Thus the magnitude of the displacement can be arbitrarily large for any $\eta < 1$ if $\alpha$ is relatively large. On the other hand, if $\alpha$ is small, the magnitude of the displacement will also be small even for small $\eta$. This simple example shows that the severity of the loss channel depends not only on the strength of the channel but also by the size of the input state, i.e., states containing many photons, will experience larger displacements than low-photon states \cite{terhal2016encoding}. This results in an interesting trade-off for GKP encoded states, since the intrinsic properties of GKP states are generally improved by considering states containing higher photon numbers. Nonetheless, it has been shown numerically that the GKP code performs well against loss and in some aspects outperforms the cat \cite{leghtas2013hardware, bergmann2016quantum, li2017cat} and binomial codes \cite{michael2016new} which are designed specifically to protect against the loss channel \cite{albert2018performance}. Furthermore, GKP states containing more photons always seem to perform better against loss than GKP states containing fewer photons \cite{albert2018performance,noh2018quantum}. 

So far, the question on how to correct loss with the GKP code in practice is usually answered by the statement that loss can be transformed into a random Gaussian displacement channel by adding a quantum-limited phase-insensitive amplification channel \cite{noh2018quantum}. The effect of the random Gaussian displacement channel is to randomly displace the input state by a magnitude that is independent of the state, and therefore, increasing the photon number of the GKP state has no negative side-effects for this channel, thus resolving the problem of potentially large displacements. However, implementing a strictly quantum-limited amplification channel is highly challenging in practice and will inevitably introduce an additional overhead of noise. Additionally, it is not clear if practically relevant GKP states are so large that the displacement associated with loss cannot be corrected without amplification. Even the ideal quantum-limited amplification channel will add noise to the state, and so it is important to consider whether this added noise contributes more logical errors than what is avoided by applying the amplification in the first place. 

Recently, a method to correct loss on GKP states without amplification was proposed by Fukui et al.\ \cite{fukui2021all} in the context of a long-distance communication protocol. In that work, the authors showed that if the loss is distributed evenly between the input mode to be corrected and the ancilla used for QEC, the combined noise on the two modes would be equivalent to a random Gaussian displacement on the input mode. Furthermore, this approach was shown to outperform strategies based on amplification in this particular setting. However, in other settings, such as in a quantum computer, one is likely unable to redistribute the loss between the encoded state and the QEC ancilla states and therefore this strategy is not universally applicable. 

In this work we quantify the performance of GKP QEC against losses both with and without added amplification. Our analysis shows that for practically relevant states, i.e.\ GKP squeezing levels below 15-20 dB, the application of an amplification channel introduces more errors than it fixes. Thus, one does not have to, and should not, worry about implementing efficient quantum-limited amplification to take advantage of the GKP encoding against loss. Our analysis is inspired and enabled by recent developments in the phase-space description of GKP states \cite{baragiola2019all,mensen2021phase,garcia2021bloch,bourassa2021fast}.

\subsection{Preliminaries}
We consider bosonic modes with position and momentum operators denoted $\hat{q}$ and $\hat{p}$, using the scaling convention of $[\hat{q},\hat{p}]=i$, corresponding to $\hbar=1$. Ideal square GKP states can be defined as superpositions of position eigenstates at integer multiples of $\sqrt{\pi}$:
\begin{subequations}\label{eq:GKPideal}
\begin{align}
    \ket{0_L} &= \sum_{s\in \mathbb{Z}} \ket{2s\sqrt{\pi}}_q,\\
    \ket{1_L} &= \sum_{s\in \mathbb{Z}} \ket{(2s+1)\sqrt{\pi}}_q.
\end{align}
\end{subequations}
These states are not normalizable and thus nonphysical. However, we can use them to mathematically construct physically valid states with the non-unitary photon-number-damping operator $e^{-\epsilon \hat{n}}$:
\begin{subequations} \label{eq:GKPeps}
\begin{align}
    \ket{0_L^\epsilon} &= e^{-\epsilon \hat{n}}\ket{0_L},\\
    \ket{1_L^\epsilon} &= e^{-\epsilon \hat{n}}\ket{1_L}, 
\end{align}
\end{subequations}
where $\hat{n}=\frac{1}{2}(\hat{q}^2+\hat{p}^2-1)$ is the number operator and $\epsilon$ denotes the strength of the damping. These states are composed of squeezed peaks in phase space of variance $\frac{1}{2}\tanh(\epsilon)$ \cite{noh2020fault,bourassa2021fast}, such that the squeezing level compared to vacuum in dB is given by $-10\log_{10}(\tanh(\epsilon))$. Furthermore, peaks far from the origin are dampened by a Gaussian envelope of width $\frac{1}{2}\tanh(\epsilon)^{-1}$, ensuring that the states have finite energy. Another subtle effect of the operator $e^{-\epsilon \hat{n}}$ is that the positions of the peaks are slightly reduced by a factor $\sech(\epsilon)\approx 1-\frac{1}{2}\epsilon^2$ in both quadratures \cite{matsuura2020equivalence}. The states of Eq.\ \eqref{eq:GKPeps} approach the ideal GKP states of Eq.\ \eqref{eq:GKPideal} in the limit of $\epsilon\rightarrow 0$.

\begin{figure}
    \centering
    \includegraphics{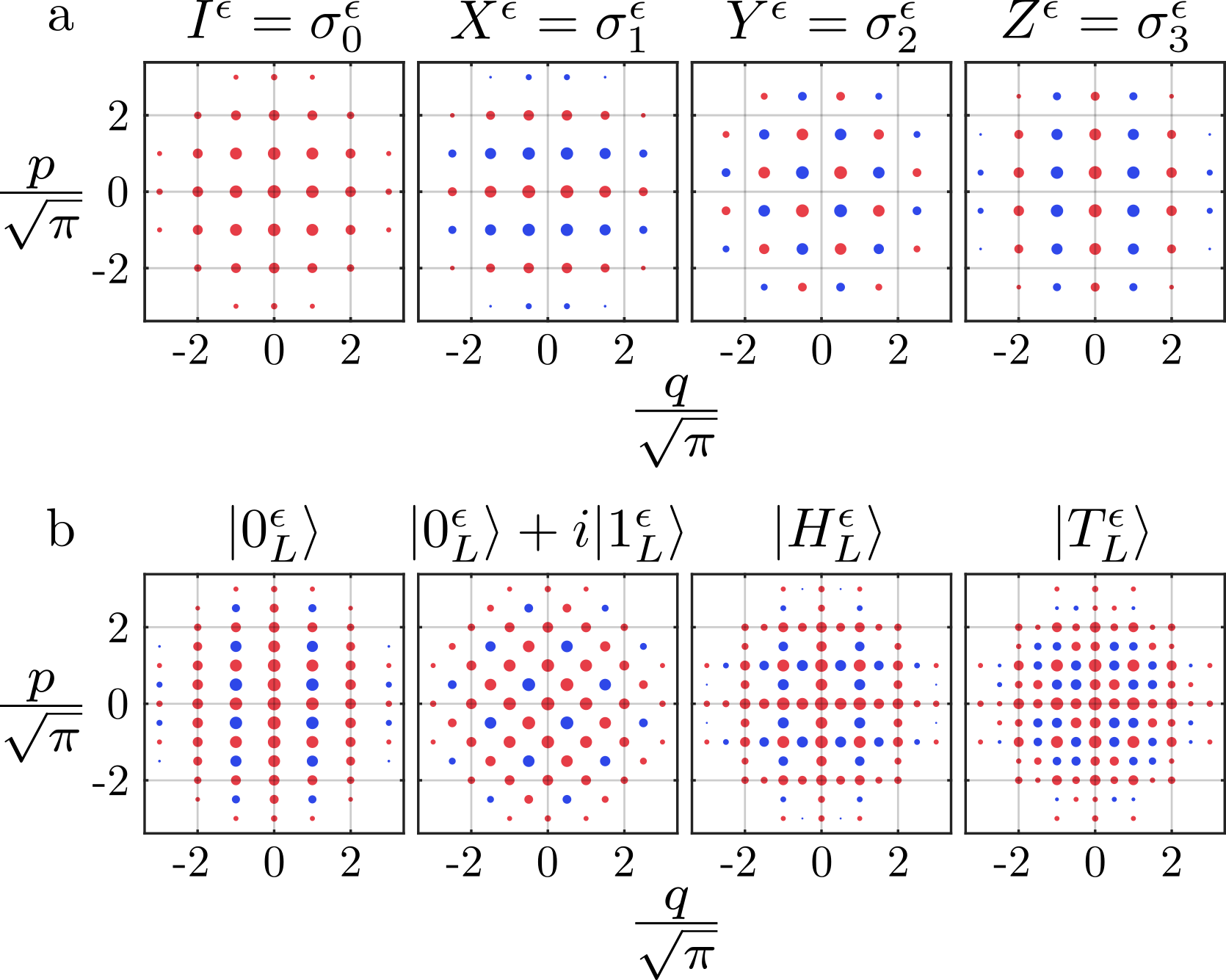}
    \caption{(a): Contours of the Wigner functions of the finite-energy GKP Pauli operators of Eq.\ \eqref{eq:GKPPauli}. Red indicates positive regions and blue indicates negative regions. (b): Wigner functions of various GKP states, namely a logical $Z$ eigenstate, a logical $Y$ eigenstate, an $H$-type magic state and a $T$-type magic state. The states are described by the GKP Pauli vectors $[0,0,1]$, $[0,1,0]$, $[\frac{1}{\sqrt{2}},0,\frac{1}{\sqrt{2}}]$ and $[\frac{1}{\sqrt{3}},\frac{1}{\sqrt{3}},\frac{1}{\sqrt{3}}]$ respectively.}
    \label{fig:Wigner}
\end{figure}

We will refer to the finite-energy GKP states simply as GKP states in the remainder of this paper, with the superscript $\epsilon$ reminding us that these are not ideal GKP states. We now use $\ket{0_L^\epsilon}$ and $\ket{1_L^\epsilon}$ to define a logical GKP identity operator
\begin{subequations}  \label{eq:GKPPauli}
\begin{equation}
I^\epsilon \equiv \sigma_0^\epsilon \equiv \ket{0_L^\epsilon}\bra{0_L^\epsilon} + \ket{1_L^\epsilon}\bra{1_L^\epsilon},
\end{equation}
as well as logical Pauli operators
\begin{align}
    X^\epsilon\equiv\sigma_1^\epsilon &\equiv \ket{1_L^\epsilon}\bra{0_L^\epsilon} + \ket{0_L^\epsilon}\bra{1_L^\epsilon},\\
    Y^\epsilon\equiv\sigma_2^\epsilon &\equiv i\ket{1_L^\epsilon}\bra{0_L^\epsilon} - i\ket{0_L^\epsilon}\bra{1_L^\epsilon},\\
    Z^\epsilon\equiv\sigma_3^\epsilon &\equiv \ket{0_L^\epsilon}\bra{0_L^\epsilon} - \ket{1_L^\epsilon}\bra{1_L^\epsilon}.
\end{align}
\end{subequations} 
Using these, states in the GKP sub-space can be written as
\begin{equation}
    \rho^{\epsilon} = \frac{1}{N}\left(\sigma_0^\epsilon + \sum_{k=1}^3a_k\sigma_k^\epsilon\right), \label{eq:GKPLogical}
\end{equation}
where $\vec{a}=[a_1,a_2,a_3]$ is a GKP Pauli vector which characterizes the state, analogous to a conventional qubit Pauli vector \cite{mensen2021phase}. The normalization factor, $N$, is given by
\begin{equation}
    N = \Tr(\sigma_0^\epsilon) + \sum_{k=1}^{3} a_{k}\Tr(\sigma_k^\epsilon).  \label{eq:Normalization}
\end{equation}
The Wigner functions of the GKP Pauli operators are shown in Fig.\ \ref{fig:Wigner}a. Since the Wigner function is linear in the density matrix, the Wigner functions of logical GKP states are constructed by simply adding together the Wigner functions of $\sigma_0^\epsilon$, $\sigma_1^\epsilon$, $\sigma_2^\epsilon$ and $\sigma_3^\epsilon$ according to the weighting given by $\vec{a}$. Some examples are shown in Fig.\ \ref{fig:Wigner}b. Note, that unlike conventional qubit Pauli operators, the GKP Pauli operators have non-zero trace (except for $\sigma_2^\epsilon$ for which $\Tr(\sigma_2^\epsilon)=0$ due to its anti-symmetric Wigner function for all $\epsilon$). As a result, the normalization factor $N$ depends slightly on $\vec{a}$ as per Eq.\ \eqref{eq:Normalization}. Still, for all $\epsilon>0$ the states defined by Eq.\ \eqref{eq:GKPLogical} are physically valid states as long as $|\vec{a}|\leq 1$. Furthermore, for small $\epsilon$ the traces of $\sigma_1^\epsilon$ and $\sigma_3^\epsilon$ vanish compared to that of $\sigma_0^\epsilon$. 

Describing GKP states in terms of their GKP Pauli vectors allows us to analyse them as qubits, despite being embedded in a larger continuous-variable Hilbert space. For example, we can define a logical fidelity between two GKP states with GKP Pauli vectors $\vec{a}$ and $\vec{b}$ as
\begin{equation}
    F_{L}(\vec{a},\vec{b}) = \frac{1}{2}\left(1 + \vec{a}\cdot\vec{b} + \sqrt{(1 - |\vec{a}|^2)(1-|\vec{b}|^2)}\right). \label{eq:fidelity}
\end{equation}
This definition is motivated by the fact that if $\vec{a}$ and $\vec{b}$ are the Pauli vectors of two $\textit{qubit}$ states $\rho_a$ and $\rho_b$, their Uhlmann fidelity, $\Tr(\sqrt{\sqrt{\rho_a}\rho_b\sqrt{\rho_a}})^2$, is given exactly by Eq.\ \eqref{eq:fidelity}. Note that the logical fidelity defined by Eq.\ \eqref{eq:fidelity} for GKP states is not the same as their Uhlmann fidelity. This is most evident in the limit of large $\epsilon$, for which both $\ket{0_L^\epsilon}$ and $\ket{1_L^\epsilon}$ converge to the vacuum state. In this regime, the Uhlmann fidelity between, e.g. $\ket{0_L^\epsilon}$ and $\ket{1_L^\epsilon}$ approaches 1, whereas their logical fidelity remains 0. By characterizing, e.g., a channel fidelity in terms of the Uhlmann fidelity, this convergence to vacuum can lead to high fidelities at large $\epsilon$, which would be misleading as GKP states do not represent useful qubits in that regime. In the regime of small $\epsilon$, which is more relevant for applications, the logical fidelity and the Uhlmann fidelity converge. 

\begin{figure}
    \centering
    \includegraphics{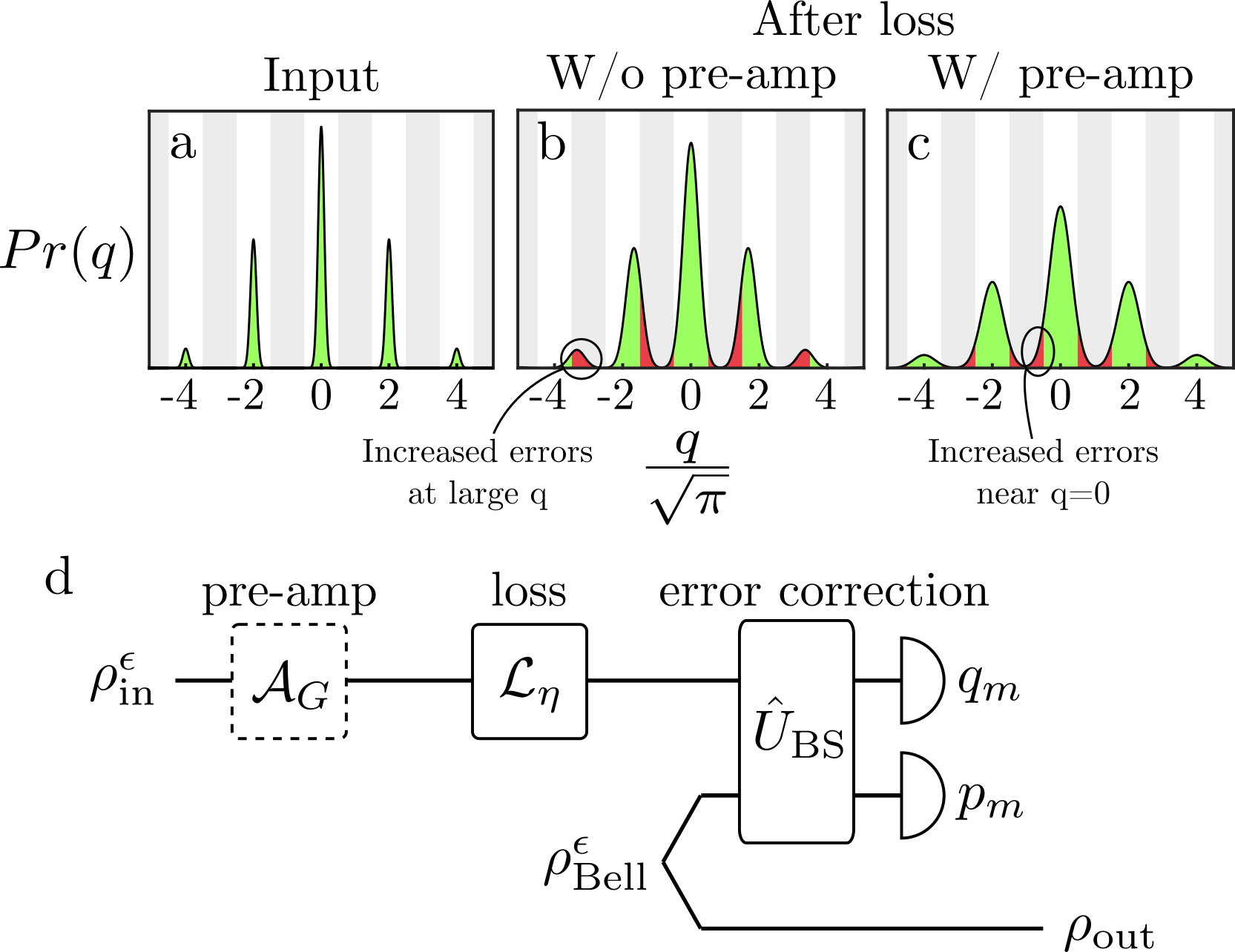}
    \caption{(a): $q$-quadrature marginal distribution of the state $\ket{0_L^\epsilon}$ with $\epsilon = 0.05$ (13 dB squeezing). (b): Marginal distribution of the state in (a) after a loss channel with $\eta = 0.7$. The red regions will be misinterpreted as a logical `1' by homodyne detection. (c): Marginal distribution of the state in (a) after amplification and loss. (d): The total channel considered in this paper, comprising noise from a loss channel $\mathcal{L}_\eta$ with transmissivity $\eta$, teleportation-based GKP QEC and a possible pre-amplification channel $\mathcal{A}_G$ with gain $G$. }
    \label{fig:Wavefunction}
\end{figure}

\subsection{Effects of loss on GKP states}
The loss channel, denoted $\mathcal{L}_\eta$, reduces the energy of an input state by a factor $\eta$ and is equivalent to mixing the input with vacuum on a beamsplitter with transmittance $\eta$. Losses have two effects on GKP states. First, the state shrinks in phase space by a factor $\sqrt{\eta}$. This shifts the peaks of the GKP state closer to the origin in phase space. Second, the variance of each peak of the GKP state is increasing towards that of the vacuum state. Specifically, the variances increase from $\frac{1}{2}\tanh(\epsilon)$ to $\frac{1}{2}\tanh(\epsilon)\eta + \frac{1-\eta}{2}$. Either of these two effects, as well as their combination, can cause parts of the wavefunction to shift by more than $\sqrt{\pi}/2$, which may lead to an erroneous bit- or phase-flip in the subsequent GKP error correction operation. This is illustrated in Fig.\ \ref{fig:Wavefunction}a and b, showing an example of a position marginal distribution of a GKP state before and after loss with the erroneous parts highlighted in red. 

The quadrature-shrinking effect of losses can be eliminated by adding a quantum-limited amplification channel, denoted $\mathcal{A}_G$, with gain $G=1/\eta$, either before the loss channel (pre-amplification) or after the loss channel (post-amplification) \cite{noh2018quantum}. However, both pre- and post-amplification comes at the cost of a further increase of the peak variances. Pre-amplification adds less noise to the state than post-amplification, and is thus considered as the reference strategy for this work. In particular, the variance after pre-amplification and loss is $\frac{1}{2}\tanh(\epsilon) + 1 - \eta$. This case is illustrated in Fig.\ \ref{fig:Wavefunction}c. As shown in the figure, the errors due to the displacement of the peaks, and in particular the peaks located further from the origin, are almost eliminated using pre-amplification. However, because of the additional broadening, more errors occur in the tail of each peak, which is most noticeable at the central peak of Fig.\ \ref{fig:Wavefunction}c. The qualitative effect of pre-amplification is thus to redistribute the errors from the outer peaks out across all peaks. However, it is not obvious whether this redistribution of errors reduces or increases the total amount of errors. Yet, we might intuitively expect that if the input state contains many peaks, i.e.\ if $\epsilon$ is very small, pre-amplification should be beneficial as peaks far from the origin otherwise experience a very large absolute displacement from the loss. 

\subsection{GKP error correction}
We now proceed by quantifying the performance of the GKP QEC protocol, comparing the pure loss channel with the amplifier-loss channel. We consider teleportation-based QEC \cite{walshe2020continuous}, which is shown in Fig. \ref{fig:Wavefunction}d. The input state after the loss channel is mixed on a 50:50 beamsplitter with one half of a GKP Bell state, $\rho_{\textrm{Bell}}^\epsilon$, and the resulting output modes are detected by two homodyne detectors measuring the conjugate quadratures, $\hat{q}$ and $\hat{p}$, respectively. The corrected input state is then recovered at the other half of the GKP Bell state by applying a corrective Pauli gate depending on the homodyne measurement outcome.

The GKP Bell state is given by
\begin{equation}
    \ket{\textrm{Bell}_L^\epsilon}=\frac{1}{\sqrt{N_\textrm{Bell}}}(\ket{0_L^\epsilon}\ket{0_L^\epsilon} + \ket{1_L^\epsilon}\ket{1_L^\epsilon}),
\end{equation}
with a corresponding density matrix
\begin{align}
    \rho_\textrm{Bell}^{\epsilon}&=\ket{\textrm{Bell}_L^\epsilon}\bra{\textrm{Bell}_L^\epsilon}\nonumber \\
    &=\frac{1}{N_\textrm{Bell}}\left(\sigma_0^\epsilon\otimes\sigma_0^\epsilon+\sigma_1^\epsilon\otimes\sigma_1^\epsilon-\sigma_2^\epsilon\otimes\sigma_2^\epsilon+\sigma_3^\epsilon\otimes\sigma_3^\epsilon\right)\nonumber \\
    &=\frac{1}{N_\textrm{Bell}}\sum_{k=0}^3 (-1)^{\delta_{k,2}} \sigma_k^\epsilon \otimes \sigma_k^\epsilon ,
\end{align}
and normalization
\begin{equation}
    N_\textrm{Bell} = \Tr(\sigma_0^\epsilon)^2 + \Tr(\sigma_1^\epsilon)^2 + \Tr(\sigma_3^\epsilon)^2,
\end{equation}
with $\delta_{i,j}$ denoting the Kronecker delta function. This Bell state can be generated by mixing two scaled GKP states, ``GKP qunaught states", on a 50:50 beamsplitter \cite{walshe2020continuous}. By mixing one half of the Bell state with the input state $\rho_0$ on the beamsplitter and obtaining homodyne measurement outcomes $q_m$ and $p_m$, the (not normalized) output state is
\begin{multline}
    \rho_\textrm{out}\\=\sum_{k=0}^3 \frac{(-1)^{\delta_{k,2}}}{N_\textrm{Bell}}\bra{q_m}\bra{p_m}\hat{U}_\textrm{BS}\left[\rho_0\otimes \sigma_k^\epsilon \right]\hat{U}_\textrm{BS}^\dagger\ket{q_m}\ket{p_m} \sigma_k^\epsilon.
\end{multline}
Defining
\begin{multline}
    \lambda_k(q_m,p_m;\rho_0)\\=\frac{(-1)^{\delta_{k,2}}}{N_\textrm{Bell}}\bra{q_m}\bra{p_m}\hat{U}_\textrm{BS}\left[\rho_0\otimes \sigma_k^\epsilon \right]\hat{U}_\textrm{BS}^\dagger\ket{q_m}\ket{p_m}, \label{eq:lambda}
\end{multline}
we can write
\begin{align}
    \rho_\textrm{out} = \lambda_0\left(\sigma_0^\epsilon + \sum_{k=1}^3\frac{\lambda_k}{\lambda_0}\sigma_k^\epsilon\right).
\end{align}
Comparing to Eq.\ \eqref{eq:GKPLogical}, the output is a GKP state described by the GKP Pauli vector, 
\begin{equation}
    \vec{a}_\textrm{out}(q_m,p_m;\rho_0) = \frac{1}{\lambda_0}[\lambda_1,\lambda_2,\lambda_3].
\end{equation}
The probability density of obtaining measurement outcome $(q_m,p_m)$ is given by:
\begin{equation}
    P(q_m,p_m;\rho_0) = \sum_{k=0}^3 \lambda_k(q_m,p_m;\rho_0)\Tr(\sigma_k^\epsilon).
\end{equation}
We use the output GKP Pauli vector to quantify the total channel comprising optional pre-amplification, loss, and error correction. Ideally, this channel should amount to a unitary Pauli rotation resulting from the teleportation protocol, with the measurement outcome dictating which Pauli rotation was applied. Thus, the ideal output GKP Pauli vector is 
\begin{equation}
    \vec{a}_\textrm{ideal}=\vec{s}(q_m,p_m) \,\circ \,\vec{a}_\textrm{in}, \label{eq:aIdeal}
\end{equation}
where `$\circ$' denotes element-wise multiplication and $\vec{s}=[s_x,s_xs_z,s_z]$ with $s_x=\pm1$ and $s_z=\pm1$ are providing sign flips according to the measurement-dependent Pauli rotation. For a given input GKP state $\rho_\textrm{in}^\epsilon$ and measurement outcomes $(q_m,p_m)$, the logical input-output fidelity of the channel is thus
\begin{equation}
    F_\textrm{in-out}(q_m,p_m;\rho_\textrm{in}^\epsilon) = F_L(\vec{a}_\textrm{out}(q_m,p_m;\rho_0),\vec{a}_\textrm{ideal}),
\end{equation}
with $\rho_0 = \mathcal{L}_\eta\left(\mathcal{A}_G\left(\rho_\textrm{in}^\epsilon\right)\right)$. Averaging over the measurement outcomes we get the mean input-output fidelity:
\begin{equation}
    \bar{F}_\textrm{in-out}(\rho_\textrm{in}^\epsilon) = \int dq_m\ dp_m\  F_\textrm{in-out}(q_m,p_m;\rho_\textrm{in}^\epsilon) P(q_m,p_m).
\end{equation}
To define a channel fidelity \cite{bowdrey2002fidelity}, we average the mean input-output fidelity over the six GKP Pauli eigenstates, i.e.\ the states with $a_\textrm{in} \in \{[\pm1,0,0],[0,\pm1,0], [0,0,\pm1]\}.$ Denoting these states as $\rho_{\pm k}^\epsilon$ with $k\in\{1,2,3\}$, the logical channel fidelity is defined as 
\begin{equation}
    F_\mathcal{C} = \frac{1}{6}\sum_{j=\pm 1,\pm 2,\pm 3}\bar{F}_\textrm{in-out}(\rho_{j}^\epsilon).
\end{equation}
In the Appendix we show how to calculate the coefficients $\lambda_k$ of Eq.\ \eqref{eq:lambda} needed to compute $F_\mathcal{C}$, for arbitrary values of $\eta$, $G$, $\epsilon$ and $\vec{a}_\textrm{in}$.

\begin{figure}
    \centering
    \includegraphics{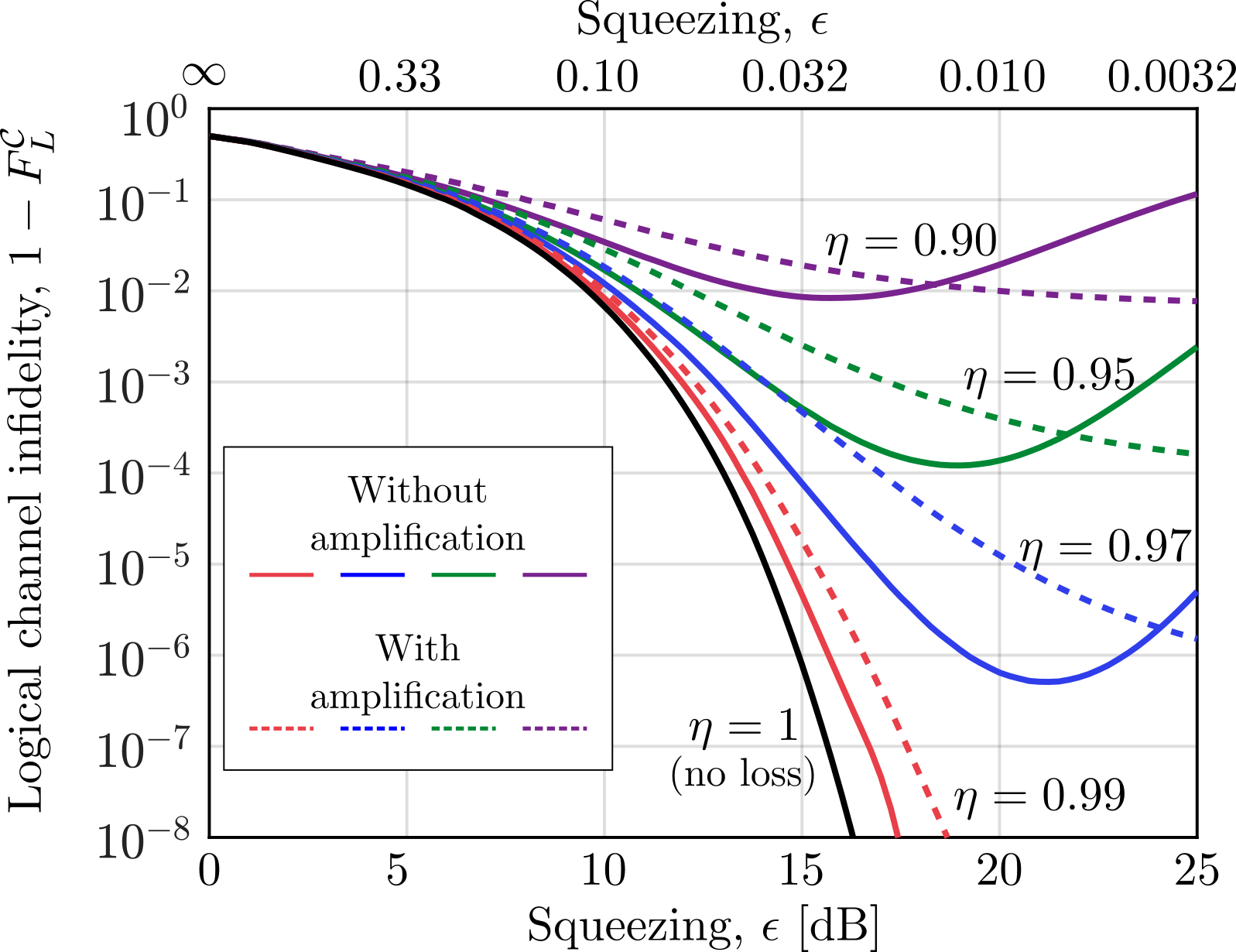}
    \caption{Logical channel infidelity for the QEC circuit depicted in Fig.\ \ref{fig:Wavefunction}d, as a function of the GKP squeezing parameter $\epsilon$ for different values of the loss channel transmissivity $\eta$. Solid lines show the results without amplification (corresponding to $G=1$) and dashed lines show the results with amplification ($G=1/\eta)$.}
    \label{fig:Results}
\end{figure}

\subsection{Results}
Fig.\ \ref{fig:Results} shows the logical channel infidelity, $1-F_\mathcal{C}$, as a function of $\epsilon$ for different amounts of loss, with (dashed lines) and without (solid lines) pre-amplification. The sign function $\vec{s}(q_m,p_m)$ of Eq.\ \eqref{eq:aIdeal} is chosen to maximize $F_\mathcal{C}$. The errors in the absence of losses (black solid line) are intrinsic to the error-correction process due to the finite squeezing of the GKP states. As the squeezing level of the GKP state increases, i.e.\ when $\epsilon$ becomes small, these errors are rapidly suppressed. When losses are considered, improving the quality of the GKP states (by reducing $\epsilon$) initially decreases the channel infidelity both with and without amplification. Importantly, the infidelity is generally lower without amplification. For very small $\epsilon$, the infidelity starts to increase with reduced $\epsilon$ in the absence of amplification. This is as expected and is due to the peak shifting effect of Fig.\ \ref{fig:Wavefunction}b, which causes errors due to the mismatch between the lattices of the GKP state after loss and the GKP Bell state used for error correction. However, pre-amplification does not beat the amplification-less strategy until the squeezing level is well above 15 dB. This is significantly larger than realistically obtainable squeezing levels, and also higher than recently estimated fault-tolerance threshold values, which are in the range from 10 to 15 dB \cite{menicucci2014fault,fukui2018high,fukui2019high,noh2020fault,noh2021low,larsen2021fault,tzitrin2021fault,walshe2021streamlined}. Thus, amplification would not realistically be required or beneficial in practical settings.

\subsection{Conclusion}
We have used newly developed phase-space methods \cite{mensen2021phase,bourassa2021fast} to model the effects of loss on finite-energy GKP states. Our analysis has shown that GKP states which have undergone loss can be error corrected without applying a loss-compensating amplification channel, provided that the GKP state does not have an unrealistically high squeezing level ($\gg 15$ dB). Furthermore, we have shown that in practically relevant regimes even ideal amplification contributes more errors than it fixes, and should therefore not be implemented in practice. This demonstrates the versatility of the GKP encoding, strengthening its candidacy as an optimal code for optical continuous-variable systems. 

\subsection{Acknowledgements}
This  project  was  supported  by  the  Danish  National Research  Foundation  through  the  Center  of  Excellence for Macroscopic Quantum States (bigQ, DNRF0142) and the European Union’s Horizon 2020 research and innovation programme CiViQ (Grant Agreement No. 820466).

\begin{widetext}

\section{Appendix}
In this Appendix we show how to calculate the output coefficients $\lambda_k$ of Eq.\ \eqref{eq:lambda}, which are used to calculate the channel fidelity of the main text. We use a phase-space representation of GKP states, in which GKP states are described as a sum of Gaussian peaks. This representation is useful, since all of the channels considered, i.e. amplification, loss, beamsplitting and homodyne detection, are Gaussian, enabling everything to be described using a Gaussian framework \cite{weedbrook2012gaussian,brask2021gaussian,bourassa2021fast}. As a result the $\lambda_k$ can be described by a sum of Gaussian functions, as we will see in the following. The key goal of the derivations in this Appendix in then to reduce the total number of Gaussian functions to a manageable level such that $\lambda$ can be computed efficiently.

The Wigner functions of the GKP Pauli states, illustrated in Fig.\ \ref{fig:Wigner}a, can be written as a sum of Gaussians \cite{bourassa2021fast,garcia2021bloch}: 
\begin{equation}
    W_k(q,p)\equiv W(q,p;\sigma_k^\epsilon) = (-1)^{\delta_{k,2}}\sum_{\boldsymbol m\in \mathcal{M}_k} c^\epsilon_{\boldsymbol m}s_k(\boldsymbol m) G_{\boldsymbol\mu_{\boldsymbol m}^\epsilon,\boldsymbol\Sigma^\epsilon}(q,p), \label{eq:GKPWigner}
\end{equation}
where
\begin{equation}
    G_{\boldsymbol\mu,\mathbf{\Sigma}}(\boldsymbol x) = \frac{1}{\sqrt{\det(\mathbf{\Sigma})(2\pi)^n}}\exp\left[-\frac{1}{2}(\boldsymbol x - \boldsymbol\mu)^T\mathbf{\Sigma}^{-1}(\boldsymbol x-\boldsymbol\mu) \right]
\end{equation}
is a multi-variable Gaussian function with mean $\boldsymbol\mu$ and covariance matrix $\mathbf{\Sigma}$. The covariance matrix and means for the approximate GKP states of Eq.\ \eqref{eq:GKPeps} are given by
\begin{equation}
    \mathbf{\Sigma}^\epsilon= \frac{1}{2}\tanh(\epsilon)\mathbb{1} \qquad \textrm{and} \qquad \boldsymbol\mu_{\boldsymbol m}^\epsilon = \sech(\epsilon)\frac{\sqrt{\pi}}{2}\boldsymbol m.
\end{equation}
where $\mathbb{1}$ is the $2\times2$ identity matrix. The vector $\boldsymbol m=\begin{bmatrix}m_1\\m_2\end{bmatrix}$ depends on the Pauli state, i.e.\ $k$, and is defined by the sets:
\begin{subequations}
\begin{align}
    \mathcal{M}_0 &= \{[m_1,m_2]^T \mid m_1,m_2 \in \mathbb{Z}, \textrm{$m_1$ is even and $m_2$ is even}\} \\ 
    \mathcal{M}_1 &= \{[m_1,m_2]^T \mid m_1,m_2 \in \mathbb{Z}, \textrm{$m_1$ is odd and $m_2$ is even}\}\\ 
    \mathcal{M}_2 &= \{[m_1,m_2]^T \mid m_1,m_2 \in \mathbb{Z}, \textrm{$m_1$ is odd and $m_2$ is odd}\}\\ 
    \mathcal{M}_3 &= \{[m_1,m_2]^T \mid m_1,m_2 \in \mathbb{Z}, \textrm{$m_1$ is even and $m_2$ is odd}\}
\end{align}
\end{subequations}
The sign function $s_k(\boldsymbol m)$ which gives rise to the negative peaks of the Wigner function, is given by:
\begin{subequations}
\begin{align}
    s_{0,\mathbf{\boldsymbol m}} &= 1\\
    s_{1,\mathbf{\boldsymbol m}} &= (-1)^{\frac{m_2}{2}}\\
    s_{2,\mathbf{\boldsymbol m}} &= (-1)^{\frac{m_1 + m_2}{2}}\\
    s_{3,\mathbf{\boldsymbol m}} &= (-1)^{\frac{m_1}{2}}
\end{align}
\end{subequations}
The additional sign flip when $k=2$, i.e.\ the factor $(-1)^{\delta_{k,2}}$ in Eq.\ \eqref{eq:GKPWigner}, could also have been included in $s_{2,\mathbf{\boldsymbol m}}$, but here we write it separately as it will simplify the following derivation. 

Finally, the weighting coefficients $c^\epsilon_{\boldsymbol m}$ are given by
\begin{equation}
    c^\epsilon_{\boldsymbol m} =  \exp\left[-\tanh(\epsilon)\frac{\pi}{4}(m_1^2+m_2^2)\right].
\end{equation}
The amplification and loss channels change the covariance matrix and mean according to
\begin{align}
    &\boldsymbol\Sigma\rightarrow G\boldsymbol\Sigma +\frac{G-1}{2}\mathbb{1}, \qquad\boldsymbol\mu \rightarrow \sqrt{G}\boldsymbol\mu \qquad\textrm{(Amplification)}\\
    &\boldsymbol\Sigma\rightarrow \eta\boldsymbol\Sigma +\frac{1-\eta}{2}\mathbb{1}, \qquad\boldsymbol\mu \rightarrow \sqrt{\eta}\boldsymbol\mu \qquad\textrm{(Loss)}
\end{align}
Thus, after pre-amplification and loss the covariance matrix is
\begin{equation}
    \boldsymbol\Sigma^\epsilon\rightarrow \tilde{\boldsymbol\Sigma}^\epsilon = \left(\frac{\eta G}{2} \tanh(\epsilon) + \eta\frac{G-1}{2} + \frac{1 - \eta}{2}\right)\mathbb{1},
\end{equation}
and the means are
\begin{equation}
    \boldsymbol\mu^\epsilon_{\boldsymbol m}\rightarrow \tilde{\boldsymbol\mu}^\epsilon_{\boldsymbol m} = \sqrt{\eta G} \sech(\epsilon)\frac{\sqrt{\pi}}{2}\boldsymbol m.
\end{equation}
Setting $G = 1/\eta$ thus leaves the means unaltered, which is the motivation for adding an amplification channel. In the following we consider arbitrary $G\geq 1$. The results of Fig.\ \ref{fig:Results} are then obtained by choosing $G=1$ and $G=1/\eta$ respectively.

The Wigner functions for the GKP Pauli states after amplification and loss are now
\begin{equation}
    \tilde{W}_k\equiv W\Big(q,p;\mathcal{L}_\eta(\mathcal{A}_G(\sigma_k^\epsilon))\Big) = (-1)^{\delta_{k,2}}\sum_{\boldsymbol m\in \mathcal{M}_k} c^\epsilon_{\boldsymbol m}s_{k,\boldsymbol m} G_{\tilde{\boldsymbol\mu}_{\boldsymbol m}^\epsilon,\tilde{\boldsymbol\Sigma}^\epsilon}(q,p).
\end{equation}
The Wigner function of an arbitrary GKP state with GKP Pauli vector $\vec{a}_\textrm{in} = [a_1,a_2,a_3]$ after amplification and loss is thus
\begin{equation}
    W\big(q,p;\mathcal{L}_\eta(\mathcal{A}_G(\rho_\textrm{in}^\epsilon))\big)=\frac{1}{N}\sum_{k=0}^3a_k \tilde{W}_k,
\end{equation}
where $a_0=1$ and $N$ is given by Eq.\ \eqref{eq:Normalization}. The Wigner function for the 3-mode state before the beamsplitter interaction of the error-correction circuit of Fig.\ \ref{fig:Wavefunction}d is then given by:
\begin{align}
    W(\boldsymbol x^{(1)},\boldsymbol x^{(2)},\boldsymbol x^{(3)}) = \sum_{k_2=0}^3\left(\frac{(-1)^{\delta_{k_2,2}}}{N N_{\textrm{Bell}}}\sum_{k_1=0}^3 a_{k_1} \tilde{W}_{k_1}(\boldsymbol x^{(1)}) W_{k_2}(\boldsymbol x^{(2)})\right) W_{k_2}(\boldsymbol x^{(3)}) \label{eq:Appendix2}
\end{align}
where $\boldsymbol x^{(i)}=(q_i,p_i)$. Here, superscript 1 refers the input mode and superscript 2 refers to the half of the GKP Bell state which is to be measured together with the input state. We now consider the term in the parentheses of the sum of Eq.\ \eqref{eq:Appendix2}, which will give us $\lambda_{k_2}$ after beamsplitting and homodyne detection \cite{bourassa2021fast}:
\begin{align}
    &\,\frac{(-1)^{\delta_{k_2,2}}}{N N_{\textrm{Bell}}}\sum_{k_1=0}^3 a_{k_1} \tilde{W}_{k_1}(\boldsymbol x^{(1)}) W_{k_2}(\boldsymbol x^{(2)}) \nonumber\\
    = &\, \frac{(-1)^{\delta_{k_1,2}}}{N N_{\textrm{Bell}}}\sum_{k_1=0}^3 a_{k_1} \sum_{\boldsymbol m^{(1)}\in \mathcal{M}_{k_1}}\sum_{\boldsymbol m^{(2)}\in \mathcal{M}_{k_2}}c^\epsilon_{\boldsymbol m^{(1)}}c^\epsilon_{\boldsymbol m^{(2)}}s_{k_1,\boldsymbol m^{(1)}}s_{k_2,\boldsymbol m^{(2)}}G_{\tilde{\boldsymbol\mu}_{\boldsymbol m^{(1)}},\tilde{\boldsymbol\Sigma}}\left(\boldsymbol x^{(1)}\right)G_{\boldsymbol\mu_{\boldsymbol m^{(2)}},\boldsymbol\Sigma}\left(\boldsymbol x^{(2)}\right) \nonumber \\
    =&\, \frac{(-1)^{\delta_{k_1,2}}}{N N_{\textrm{Bell}}}\sum_{k_1=0}^3 a_{k_1} \sum_{\boldsymbol m^{(1)}\in \mathcal{M}_{k_1}}\sum_{\boldsymbol m^{(2)}\in \mathcal{M}_{k_2}}c^\epsilon_{\boldsymbol m^{(1)}}c^\epsilon_{\boldsymbol m^{(2)}}s_{k_1,\boldsymbol m^{(1)}}s_{k_2,\boldsymbol m^{(2)}}G_{\tilde{\boldsymbol \mu}_{\boldsymbol m^{(1)}}\oplus \boldsymbol \mu_{\boldsymbol m^{(2)}},\tilde{\boldsymbol \Sigma} \oplus \boldsymbol \Sigma}\left(\boldsymbol x^{(1)},\boldsymbol x^{(2)}\right) \label{eq:Appendix1}
\end{align}
The beamsplitter is described by the symplectic matrix $S=\frac{1}{\sqrt{2}}\begin{bmatrix}\mathbb{1}& \mathbb{1}\\-\mathbb{1} & \mathbb{1}\end{bmatrix}$, which transforms the covariance and mean of each Gaussian component of the Wigner function according to 
\begin{align}
    \tilde{\boldsymbol \Sigma}\oplus \boldsymbol \Sigma \rightarrow S\left( \tilde{\boldsymbol \Sigma}\oplus \boldsymbol \Sigma \right)S^T = \frac{1}{2}\begin{bmatrix}\tilde{\boldsymbol \Sigma} + \boldsymbol \Sigma & -\tilde{\boldsymbol \Sigma} + \boldsymbol\Sigma \\ -\tilde{\boldsymbol \Sigma} + \boldsymbol \Sigma& \tilde{\boldsymbol \Sigma} + \boldsymbol \Sigma\end{bmatrix} \label{eq:covariance}\\
    \tilde{\boldsymbol \mu}_{\boldsymbol m^{(1)}}\oplus \boldsymbol \mu_{\boldsymbol m^{(2)}} \rightarrow S \left(\tilde{\boldsymbol \mu}_{\boldsymbol m^{(1)}}\oplus \boldsymbol \mu_{\boldsymbol m^{(2)}}\right)=\frac{1}{\sqrt{2}}\begin{bmatrix} \tilde{\boldsymbol \mu}_{\boldsymbol m^{(1)}} + \boldsymbol \mu_{\boldsymbol m^{(2)}}\\ -\tilde{\boldsymbol \mu}_{\boldsymbol m^{(1)}} + \boldsymbol \mu_{\boldsymbol m^{(2)}}\end{bmatrix}
\end{align}
The homodyne detectors then measure the $q$-quadrature of mode 1 and $p$-quadrature of mode 2 with measurement results $q_m$ and $p_m$ respectively, and the resulting transformation of Eq.\ \eqref{eq:Appendix1} gives the coefficient $\lambda_k(q_m,p_m;\rho_\textrm{in}^\epsilon)$. As the $q_1p_2$-element of the covariance matrix (Eq.\ \eqref{eq:covariance}) is $0$, the Gaussian factorizes, and Eq.\ \eqref{eq:Appendix1} is transformed as
\begin{align}
    &\rightarrow \frac{(-1)^{\delta_{k_1,2}}}{N N_{\textrm{Bell}}}\sum_{k_1=0}^3 a_{k_1} \sum_{\boldsymbol m^{(1)}\in \mathcal{M}_{k_1}}\sum_{\boldsymbol m^{(2)}\in \mathcal{M}_{k_2}}c^\epsilon_{\boldsymbol m^{(1)}}c^\epsilon_{\boldsymbol m^{(2)}}s_{k_1,\boldsymbol m^{(1)}}s_{k_2,\boldsymbol m^{(2)}}G_{\mu'_{\left(m_1^{(1)},m_1^{(2)}\right)},\Sigma'}(q_m)G_{\mu'_{\left(-m_2^{(1)},m_2^{(2)}\right)},\Sigma'}(p_m) \nonumber \\
    &= \lambda_{k_2}\big(q_m,p_m;\mathcal{L}_\eta(\mathcal{A}_G(\rho_\textrm{in}^\epsilon))\big),
\end{align}
where $\Sigma'$ is the upper-left element of $\frac{1}{2}(\tilde{\boldsymbol \Sigma} + \boldsymbol \Sigma)$:
\begin{equation}
    \Sigma'(\eta,G,\epsilon) = \frac{1}{4}\bigg(\tanh(\epsilon)(1 + \eta G) + \eta(G-1) + 1 - \eta\bigg),
\end{equation} 
and
\begin{equation}
    \mu_{\boldsymbol m}' = \sech(\epsilon)\frac{\sqrt{\pi}}{2\sqrt{2}}\left(\sqrt{\eta G}m_1 + m_2\right).
\end{equation}

To factorise the sums we define $\boldsymbol n^{(1)} =\left(n_1^{(1)},n_2^{(1)}\right)= \left(m_1^{(1)},m_1^{(2)}\right)$ and $\boldsymbol n^{(2)}=\left(n_1^{(2)},n_2^{(2)}\right) = \left(m_2^{(1)},m_2^{(2)}\right)$. Thus $m$ and $n$ are related through $m_i^{(j)} = n_j^{(i)}$. The sets of the summations are correspondingly changed, i.e. $\boldsymbol m^{(1)}\in\mathcal{M}_{k_1}$ and $\boldsymbol m^{(2)}\in\mathcal{M}_{k_2}$ become $\boldsymbol n^{(1)}\in\mathcal{M}_{l_1}$ and $\boldsymbol n^{(2)}\in\mathcal{M}_{l_2}$ with $l_1$ and $l_2$ depending on $k_1$ and $k_2$. For example, for $(k_1,k_2) = (0,2)$ the elements $(\boldsymbol m^{(1)},\boldsymbol m^{(2)}) = (m_1^{(1)},m_2^{(1)},m_1^{(2)},m_2^{(2)}) = (n_1^{(1)},n_1^{(2)},n_2^{(1)},n_2^{(2)})$ are (even, even, odd, odd)  so $(\boldsymbol n^{(1)},\boldsymbol n^{(2)}) = (n_1^{(1)},n_2^{(1)},n_1^{(2)},n_2^{(2)})$ is (even, odd, even, odd) corresponding to $(l_1,l_2) = (3,3)$. The general transformation rules for $(k_1,k_2)\rightarrow(l_1,l_2)$ are:
\begin{align}
    (0,0) &\rightarrow (0,0) & (1,0)&\rightarrow(1,0) & (2,0)&\rightarrow(1,1) & (3,0)&\rightarrow(0,1) \nonumber\\
    (0,1) &\rightarrow (3,0) & (1,1)&\rightarrow(2,0) & (2,1)&\rightarrow(2,1) & (3,1) &\rightarrow(3,1)\nonumber \\
    (0,2)&\rightarrow (3,3) & (1,2)&\rightarrow(2,3) & (2,2)&\rightarrow(2,2) & (3,2)& \rightarrow(3,2)\nonumber \\
    (0,3)&\rightarrow (0,3) & (1,3)&\rightarrow(1,3) & (2,3)&\rightarrow(1,2) & (3,3)&\rightarrow(0,2). \label{eq:transformation1}
\end{align}
With these new summation indices we get
\begin{equation}
     \lambda_{k_2}= \frac{(-1)^{\delta_{k_1,2}}}{N N_{\textrm{Bell}}}\sum_{k_1=1}^4 a_{k_1}\sum_{\boldsymbol n^{(1)}\in\mathcal{M}_{l_1}}c^\epsilon_{\boldsymbol n^{(1)}}G_{\mu'_{\boldsymbol n^{(1)}},\Sigma'}(q_m) \sum_{\boldsymbol n^{(2)}\in\mathcal{M}_{l_2}}c^{\epsilon}_{\boldsymbol n^{(2)}}s_{k_1,\boldsymbol m^{(1)}}s_{k_2,\boldsymbol m^{(2)}} G_{\mu'_{\left(-n^{(2)}_1,n^{(2)}_2\right)},\Sigma'}(p_m)
\end{equation}  
where we have also used the fact that $c^\epsilon_{\boldsymbol m^{(1)}}c^\epsilon_{\boldsymbol m^{(2)}}=c^\epsilon_{\boldsymbol n^{(1)}}c^\epsilon_{\boldsymbol n^{(2)}}$. The sign functions $s_{k_1,\boldsymbol m^{(1)}}$ and $s_{k_2,\boldsymbol m^{(2)}}$ can also be factorized in terms of $\boldsymbol n^{(1)}$ and $\boldsymbol n^{(2)}$ with a suitable change of indices. For example, for $(k_1,k_2) = (0,2)$ we get $s_{k_{1},\boldsymbol m^{(1)}}s_{k_{2},\boldsymbol m^{(2)}}=1\cdot(-1)^{(m_1^{(2)}+m_2^{(2)})/2}=(-1)^{n_2^{(1)}/2}(-1)^{n_2^{(2)}/2}=s_{l'_1,\boldsymbol n^{(1)}}s_{l'_2,\boldsymbol n^{(2)}}$ with $(l_1',l_2') = (1,1)$. The general transformation rules for $(k_1,k_2)\rightarrow(l_1',l_2')$ are:
\begin{align}
    (0,0) &\rightarrow (0,0) & (1,0)&\rightarrow(0,3) & (2,0)&\rightarrow(3,3) & (3,0)&\rightarrow(3,0) \nonumber\\
    (0,1) &\rightarrow (0,1) & (1,1)&\rightarrow(0,2) & (2,1)&\rightarrow(3,2) & (3,1) &\rightarrow(3,1)\nonumber \\
    (0,2)&\rightarrow (1,1) & (1,2)&\rightarrow(1,2) & (2,2)&\rightarrow(2,2) & (3,2)& \rightarrow(2,1)\nonumber \\
    (0,3)&\rightarrow (1,0) & (1,3)&\rightarrow(1,3) & (2,3)&\rightarrow(2,3) & (3,3)&\rightarrow(2,0). \label{eq:transformation2}
\end{align}

We thus have
\begin{equation}
   \lambda_{k_2} = \frac{(-1)^{\delta_{k_1,2}}}{N N_{\textrm{Bell}}}\sum_{k_1=1}^4 a_{k_1}\sum_{\boldsymbol n^{(1)}\in\mathcal{M}_{l_1}}c_{\boldsymbol n^{(1)}}s_{l'_1,\boldsymbol n^{(1)}}G_{\mu'_{\boldsymbol n^{(1)}},\Sigma'}(q_m) \sum_{\boldsymbol n^{(2)}\in\mathcal{M}_{l_2}}c_{\boldsymbol n^{(2)}}s_{l'_2,\boldsymbol n^{(2)}} G_{\mu'_{\left(-n^{(2)}_1,n^{(2)}_2\right)},\Sigma'}(p_m) 
    \end{equation}
Finally, we flip the sign of $n_1^{(2)}$ in the last sum. $c_{\boldsymbol n^{(2)}}$ does not depend on the sign of the elements of $\boldsymbol n^{(2)}$. For $s_{l_2',\boldsymbol n^{(2)}}$, changing the sign of $n_1^{(2)}$ yields an additional minus sign when $s_{l_2',n^{(2)}}$ depends on $n_1^{(2)}$, i.e.\ when $l_2'\in [2,3]$, and $n_1^{(2)}$ is odd, i.e.\ $l_2\in [1,2]$. From the transformation rules of Eqs.\ \eqref{eq:transformation1} and \eqref{eq:transformation2}, this occurs if and only if $k_1=2$, conveniently cancelling out the factor $(-1)^{\delta_{k_1,2}}$. Thus we get:
\begin{align}
\lambda_{k_2}=\frac{1}{N N_{\textrm{Bell}}}\sum_{k_1=1}^4 a_{k_1}\sum_{\boldsymbol n^{(1)}\in\mathcal{M}_{l_1}}c_{\boldsymbol n^{(1)}}s_{l'_1,\boldsymbol n^{(1)}}G_{\mu'_{\boldsymbol n^{(1)}},\Sigma'}(q_m) \sum_{\boldsymbol n^{(2)}\in\mathcal{M}_{l_2}}c_{\boldsymbol n^{(2)}}s_{l'_2,\boldsymbol n^{(2)}} G_{\mu'_{\boldsymbol n^{(2)}},\Sigma'}(p_m). 
\end{align}
Defining
\begin{equation}
g_{l,l'}(x) \equiv \sum_{\boldsymbol n\in\mathcal{M}_l}c_{\boldsymbol n} s_{l',\boldsymbol n}G_{\mu'_{\boldsymbol n},\Sigma'}(x),
\end{equation}
we can write this as
\begin{equation}
    \lambda_{k_2}= \frac{1}{N N_{\textrm{Bell}}}\sum_{k_1=1}^4 a_{k_1} g_{l_1,l_1'}(q_m)  g_{l_2,l_2'}(p_m).
\end{equation}
The functions $g_{l,l'}(x)$ can be calculated numerically by cutting off the set $\mathcal{M}_l$ when $c_{\boldsymbol n} = \exp\left[-\tanh(\epsilon)\frac{\pi}{4}|\boldsymbol n|^2\right]$ is sufficiently small. For the results of Fig.\ \ref{fig:Results} we chose $c_{\boldsymbol n}>\exp(-23)\approx 10^{-10} \Leftrightarrow |\boldsymbol n|<\sqrt{23\times \frac{4}{\pi}\tanh(\epsilon)^{-1}}$

\end{widetext}
\section{References}
\bibliography{References} 

\end{document}